\documentclass{epl}

\usepackage{graphicx,bm,amssymb,amsmath}

\def\be{\begin{equation}}
\def\fin{\end{equation}}

\def\T{{\sf T\kern-.45em T}}
\def\C{\kern.1em{\raise.47ex\hbox{$\scriptscriptstyle |$}}
             \kern-.40em{\sf C}}

\title{Spontaneous flow transition in active polar gels}
\shorttitle{spontaneous flow transition}
\author{ R.Voituriez$^{1}$, J.F. Joanny$^1$ and J. Prost$^{1,2}$ }

\institute{
  \inst{1} Physicochimie Curie (CNRS-UMR168), Institut Curie, Section 
de Recherche, 
26 rue d'Ulm 75248 Paris Cedex 05 France\\
  \inst{2} E.S.P.C.I, 10 rue Vauquelin, 75231 Paris Cedex 05, France
}

\begin{document}

\maketitle

\begin{abstract}
We study theoretically the effects of confinement on active polar
gels such as the actin network of eukaryotic cells. Using 
generalized hydrodynamics equations 
derived for active gels, we predict, in the case of quasi one-dimensional geometry, 
a spontaneous flow transition from a homogeneously polarized immobile 
state for small thicknesses, to a perturbed flowing state 
for larger thicknesses. The transition is not driven by an 
external field but by the activity of the system. We suggest several possible experimental realizations.

\end{abstract}
\section{Introduction and statement of the problem}
Active materials are a challenging class of systems
driven out of equilibrium by an internal or an external energy
source. Many  examples of active systems are provided 
by the biological world such as self--propelled particle assemblies 
in bacterial colonies, or the membrane or the cytoskeleton
of eukaryotic cells~\cite{albe02}. The cell cytoskeleton
is a complex network of long filamentary proteins (mostly F-actin,
microtubules and intermediate filaments) interacting with a variety of smaller
proteins~\cite{howa01} which can, among other things, crosslink or cap the filaments. 
A well studied class of proteins interacting with actin and microtubules are   motor proteins, myosin, kinesin or dyneins.  These proteins use the chemical
energy of Adenosinetriphosphate (ATP) hydrolysis to "walk" along the filaments,
and exert stresses that deform the filament network~\cite{taki91,nede97}. 
The active properties of the cytoskeleton play a crucial role in most  for 
cell functions such as intracellular transport, motility and cell division.

Many efforts towards  understanding  the mechanical properties of the cytoskeleton  have 
focused on the description of its passive visco-elastic properties which are  
well understood in terms of a gel built by cross-linked semi-flexible polymers 
\cite{head03,wilh03}. More recently, Kruse et al. \cite{krus04,krus05} have proposed 
a generalized hydrodynamic theory based on conservation
laws and symmetry considerations, to describe macroscopically {\it active} polar gels.  
A typical example is given by the network of actin cytoskeletal
filaments  in the presence of myosin II motor proteins which generate active processes 
by  hydrolyzing ATP. Since cytoskeletal filaments are structurally polar (with a + and - end), each filament locally defines
a unit vector. The filamental structure gives rise on large scales to a macroscopic 
polarity if the filaments are on average aligned. 

Experiments, numerical simulations and analytical descriptions, 
have shown that the cell cytoskeleton has a rich and complex 
mechanical behavior
 \cite{taki91,nede97,krus00,krus01,lee01,kim03,live03,seki91}.
In particular, self-organized  patterns, including asters, vortices, and 
rotating spirals have been observed as a function of
motor and ATP concentrations in a two-dimensional
geometry~\cite{nede97}, and have been recently reproduced 
theoretically \cite{krus04,menon}. 

It is well known that boundary effects can play a very important role in the formation of  self-organized patterns. Here, we use the generalized hydrodynamic description 
of Ref. \cite{krus04,krus05} to study analytically the effect of confinement 
of active polar gels confined between two parallel surfaces for 
various kinds of boundary conditions 
(free, no--slip, mixed and active).  
We predict a "Frederiks--like" flow transition 
from a homogeneously polarized immobile state for small thicknesses, 
to a inhomogeneous flowing state for larger thicknesses. This transition is reminiscent of the classical Frederiks transition in thin nematic liquid-crystal films since the system switches from a homogeneous to a non homogeneous polarization state. However there are two important differences, the transition in active polar gels does not require any external field and the non homogeneous active state is mobile. The critical length at which the transition occurs is monitored by the active stress proportional to the 
ATP/ADP chemical potential difference $\Delta \mu$. An experimental study of 
this spontaneousflow transition could give access to the active parameters 
of the actin-myosin cytoskeleton.

\begin{figure}
\begin{center}\scalebox{0.32}{
\includegraphics{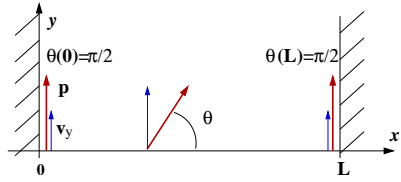}}
\caption{\label{figure:fig1} quasi--1d confined geometry}\end{center}
\end{figure}

We now expose briefly the model in 2 dimensions, following Ref.\cite{krus04,krus05}. 
The network of actin filaments has a macroscopic polarity described by a 
unit vector polarization field ${\bf p}=(\cos \theta,\sin\theta)$ (see Fig.\ref{figure:fig1}). The associated 
polarization free energy is given by the standard expression for a polar liquid crystal \cite{dege93} :
\begin{equation}
F=\int dxdy\left[ \frac{K_1}{2}(\nabla\cdot {\bf  p} )^2 +\frac{K_3}{2}(\nabla\times{\bf  p})^2 -
\frac{1}{2}h_\parallel {\bf p}^2 \right]
\end{equation}
where $K_1=K$ and $K_3$ are the splay and bend
elastic moduli. Note that there is no twist free energy in 2 dimensions, and that we do not write here the spontaneous splay term $k\nabla\cdot {\bf  p}$ allowed by polar  symmetry. This term is equivalent to a surface free energy of the form $k({\bf p}\cdot{\bf n})$ (${\bf n}$ being the unit vector normal to the surface). In the strong anchoring limit, this term is small compared to the anchoring energy that could be written as $k'({\bf p}\cdot{\bf n})^2$ with a large  energy $k'\to\infty$. In this strong anchoring limit, the elastic free energy is the same as that of a classical nematic liquid crystal and specific polarity effects only show up in the dynamical equations \ref{uab}, \ref{eq:dpdt} below. In the case of weak anchoring, the spontaneous splay could become dominant;  as it does not allow for a state of uniform polarization, it would exclude the transition described at length in this paper. 
The Lagrange multiplier $h_{||}$ is introduced in order 
to satisfy the constraint ${\bf p}^2=1$. The molecular field, conjugate to the polarization is given by 
$h_\alpha=-\delta F/\delta p_\alpha$;
in the following, we use  its parallel and perpendicular coordinates $(h_\parallel,h_\perp)$ in the local frame linked to the polarization ${\bf p}$.

The gel motion is described by the velocity field ${\bf v}$ 
or the strain rate tensor $u_{\alpha\beta}=(\partial_{\alpha}v_\beta+\partial_{\beta}v_\alpha)/2$. 
The vorticity tensor $\omega_{\alpha\beta}=(\partial_{\alpha}v_\beta-\partial_{\beta}v_\alpha)/2$ 
is its antisymmetric counterpart. We assume here that the gel is incompressible.
The gel is driven out of equilibrium by a constant 
chemical potential difference $\Delta \mu$ between ATP and its hydrolysis products. We consider here a visco-elastic gel which has a liquid behavior at long time scales. In a steady state, the gel behaves as a Newtonian liquid and its elasticity is irrelevant. The case of active nematic elastomers would lead to a different behaviour not described by our theory, as the classical Frederiks transition in these materials is very different from that of usual  nematic liquids \cite{warner}. 
The linear generalized 
hydrodynamics equations for an incompressible active polar gel, at long time scales, read \cite{krus04,krus05}: 
\begin{eqnarray}
2\eta u_{\alpha\beta} & = & \sigma_{\alpha\beta} 
+ \zeta\Delta\mu p_\alpha p_\beta  -\frac{\nu}{2}(p_\alpha h_\beta+p_\beta h_\alpha)
+\frac{1}{2}(p_\alpha h_\beta-p_\beta h_\alpha ) +{\bar \zeta}\Delta\mu\delta_{\alpha\beta}
 \label{uab}\\
\frac{D p_\alpha}{D t}  &=&  \frac{1}{\gamma} h_\alpha 
+ \lambda p_\alpha \Delta\mu - \nu u_{\alpha\beta}p_\beta \label{eq:dpdt}
\end{eqnarray}
where we have used the corotational time derivative of the vector $p_\alpha$,
$\frac{D}{Dt}p_\alpha=\frac{\partial p_\alpha}{\partial t}+(v_\gamma\partial_\gamma)p_\alpha
+\omega_{\alpha\beta} p_\beta$. The full derivation of these constitutive equations is given  in \cite{krus05} and will not be discussed here. 
The rotational viscosity $\gamma$ and the coupling constant between
 flow and polarization $ \nu$  are standard liquid crystal 
parameters \cite{dege93}. The active contributions to the mechanical stress 
and to the rate of variation of the polarization are proportional to $\Delta \mu$ and 
are characterized by the coefficients $\zeta, {\bar \zeta}$ and $\lambda$. We limit here the theory to linear order in the activity and we neglect the geometric non-linearities introduced in \cite{krus04,krus05}. The diagonal active term proportional to $\Delta\mu\delta_{\alpha,\beta}$ contribute to the absolute value of the pressure in an incompressible gel.
 
This set of constitutive equations is completed by the force balance equation:
$\partial_\alpha (\sigma_{\alpha\beta}-\Pi\delta_{\alpha\beta})=0 $;
locally, there are two forces acting on the gel,
the deviatory stress tensor $\sigma_{\alpha\beta}$ and the pressure $\Pi$ that insures the incompressibility of the gel.

\section{Spontaneous flow transition}
We now look for the stationnary states of an  active film in a confined geometry:
we assume that the system is translationally invariant along the $y$ direction 
and that the gel is confined between  $x=0$ and $x=L$. We consider several 
hydrodynamic boundary conditions. As discussed above, in the strong anchoring limit the polarization orientation is strictly defined on the 
confining surfaces. For simplicity, we only 
consider here the planar anchoring  where the polarization at the confining surfaces is parallel to 
the $y$ direction, $\theta=\frac{\pi}2$. Similar results are obtained for an homeotropic anchoring, 
$\theta =0$.

The symmetry of the problem implies  that $v_x={\rm const}=0$, and that the
 transverse component of the stress tensor  $\sigma_{xy}$ is constant (the force balance imposes that $\partial_x \sigma_{xy}=0$). 
The perpendicular component of the molecular field is $h_\perp=K\partial_x^2\theta$. This relation is exact if $K_1=K_3=K$, and only valid for small values of $\theta -\pi/2$ if  $K_1=K \neq K_3$.  In this slab geometry, the constitutive equations for the gel motion are rewritten as 
\begin{eqnarray}
2\eta u= \frac{\sigma_{yx}+\sigma_{xy}}{2}+\frac{\zeta\Delta\mu}{2}\sin2\theta-
\frac{\nu}{2}(h_\parallel\sin2\theta+h_\perp\cos2\theta);\qquad 
\frac{\sigma_{xy}-\sigma_{yx}}{2}+\frac{h_\perp}{2} =0
\label{u}
\end{eqnarray}
where for simplicity we have denoted $u\equiv u_{xy}$.
In turn, the polarization equation \ref{eq:dpdt} gives
\begin{eqnarray}
u\nu\sin2\theta =\frac{h_\parallel}{\gamma}+\lambda\Delta\mu;\qquad
u(\nu\cos2\theta-1) = \frac{h_\perp}{\gamma}
\label{p}
\end{eqnarray}

\section{Hydrodynamic free boundary conditions}
We first consider a free standing film where the gel slides freely on the confining surfaces.  The transverse stress $\sigma_{xy}(x)$ vanishes since
$\sigma_{xy}(x=0)=\sigma_{xy}(x=L)=0$.  With this boundary condition, 
equations \ref{u},\ref{p} can be recast into a differential equation for the angle  
$\theta$ giving the polarization orientation:
\begin{equation}
\partial_x^2\theta=\frac{{\tilde \zeta}\Delta\mu\sin2\theta(\nu\cos2\theta-1)}
{K[4\frac{\eta}{\gamma}+\nu^2-2\nu\cos2\theta+1]}=\Phi_f(\theta)
\end{equation}
with ${\tilde \zeta}=\zeta +\nu\gamma\lambda$. Here we assume ${\tilde \zeta}\Delta\mu<0$, which has been shown to be the case generating self--motion of point--like defect in Ref. \cite{krus04}. A full understanding of the microscopic origin of the penomenological couplings $\zeta, \lambda$ is still missing. However, it is known experimentally \cite{thou} that for the actin-myosin gels forming the cell cytoskeleton, the stress is contractile corresponding to $\zeta\Delta\mu<0$ (a contractile effect requires a positive stress $\sigma_{\alpha\beta}$, all other terms being zero in Eq.\ref{uab}). The contractility has also been reproduced by microscopic models \cite{live03}. Eq. \ref{eq:dpdt} suggests that $\lambda\Delta\mu$ must be positive in order to account for the self--alignement effects such as the zipping effects observed in \cite{uhde04}.  
The effective potential $V_f$ associated to the force $\Phi_f(\theta)$,
$V_f=-\int\Phi_f(\theta)d\theta$,
can be expanded around the value imposed by the anchoring on 
the confining surfaces, $\theta=\pi/2+\epsilon$:
\begin{equation}\label{Vdev}
V_f(\theta)=V_f(\pi/2)-\frac{{\tilde \zeta}\Delta\mu(\nu+1)}
{K(4\eta/\gamma+(\nu+1)^2)}\epsilon^2+o(\epsilon^2)=
V_f(\pi/2)+\frac{\epsilon^2}{\ell^2}+o(\epsilon^2)
\end{equation}
Following the standard argument for nematic liquid crystals~\cite{dege93}, 
if the length  $\ell$ defined by Eq.\ref{Vdev} is real (i.e. if the effective active stress is actually contractile ${\tilde \zeta}<0$ ), 
the system exhibits a continuous spontaneous flow transition at constant activity for a critical size  $L_c=\ell\pi/\sqrt{2}$, or equivalently at constant thickness for a critical activity
\begin{equation}
\Delta\mu_c=\frac{\pi^2K(4\eta/\gamma+(\nu+1)^2)}{-2L^2{\tilde \zeta}(\nu+1)}
\end{equation}
For $L<L_c$ (or $\Delta\mu<\Delta\mu_c$), the system is dominated by anchoring effects and  the orientation of the polarization is constant $\theta(x)=\pi/2$,  
whereas for $L>L_c$ ( $\Delta\mu>\Delta\mu_c$) the polarization is tilted in the bulk down to  
a minimum angle $\theta_{min}=\theta(L/2)<\pi/2$.
We recall that close to threshold, 
$L\sim L_C$, the tilt amplitude is $\epsilon(x)=\epsilon_m\sin (x\pi/L)$ 
with $\epsilon_m\propto\sqrt{L-L_c}$ or $\epsilon_m\propto\sqrt{\Delta\mu-\Delta\mu_c}$. 

The flow field can be calculated at linear order in $\epsilon_m$ from the velocity gradient
$u=({\tilde \zeta}\Delta\mu\sin2\theta)/(4\eta+\gamma(\nu^2-2\nu\cos2\theta+1))$.
 We choose here the integration constant (the velocity at the midplane) in such a way 
that the total net flow vanishes (see Fig\ref{fig3}a): 

\begin{equation}
v_y=\frac{4L{\tilde \zeta}\Delta\mu\epsilon_m\cos(x\pi/L)}{\pi(4\eta+\gamma(\nu+1)^2)}.
\end{equation}

In the case of an infinite system along the $x$ direction, the polarization reaches the minimum tilt angle defined by $\cos 2\theta_{min} =1/\nu$ when $\vert \nu \vert >1$. If $\vert \nu \vert <1$ the angle decreases to zero. Note that when $\nu =0$ the problem can be mapped exactly onto the original Frederiks problem for any angle value and $K_1 \neq K_3$. As opposed to a liquid crystal in an externally imposed shear flow, there is no flow tumbling here because the transverse stress $\sigma_{xy}$ vanishes. One can check directly by plotting the effective potential $V_f(\theta)$ that the state $\theta=0$ is stable.

\section{Hydrodynamic no-slip boundary conditions}

For solid confining surfaces when the gel does not slide on the walls,  $v_y(x=0)=v_y(x=L)=0$. The equation for the polarization angle $\theta$ is in this case:
\begin{equation}\label{ns}
\partial_x^2\theta=\frac{({\tilde \zeta}\Delta\mu\sin2\theta+2\sigma_{xy})
(\nu\cos2\theta-1)}{K[4\eta/\gamma+\nu^2-2\nu\cos2\theta+1]}=\Phi_{ns}(\theta)
\end{equation}

At linear order in the polarization tilt, the polarization equation reads
\begin{equation}
\partial_x^2\epsilon=\frac{2(\nu+1)}{K[4\eta/\gamma+(\nu+1)^2]}
({\tilde \zeta}\Delta\mu\epsilon -\sigma_{xy})
\end{equation}
and the polarization angle is given by 
\begin{equation}
\epsilon=\frac{\sigma_{xy}}{{\tilde \zeta}\Delta\mu}\left(1-\cos(\pi x/L_c)-\tan(\frac{\pi L}{2L_c})\sin(\pi x/L_c)\right)
\label{tilt}
\end{equation}

If $\sigma_{xy}$ were different from $0$, there would be no spontaneous flow 
transition. However, we can show that in the low activity regime  $\sigma_{xy}=0$. Indeed when the activity $\Delta\mu$ is small, the tilt angle $\epsilon$ is small. One can then directly check using equation \ref{tilt} that the velocity gradient does not change sign for any constant value of $\sigma_{xy}\neq 0$. This is incompatible with the no-slip boundary conditions and necessarily $\sigma_{xy}= 0$. In turn the polarization remains uniform. For larger $\Delta \mu$  a spontaneous flow transition is reached.

\begin{figure}[h]
\begin{center}
\scalebox{0.15}{
\includegraphics{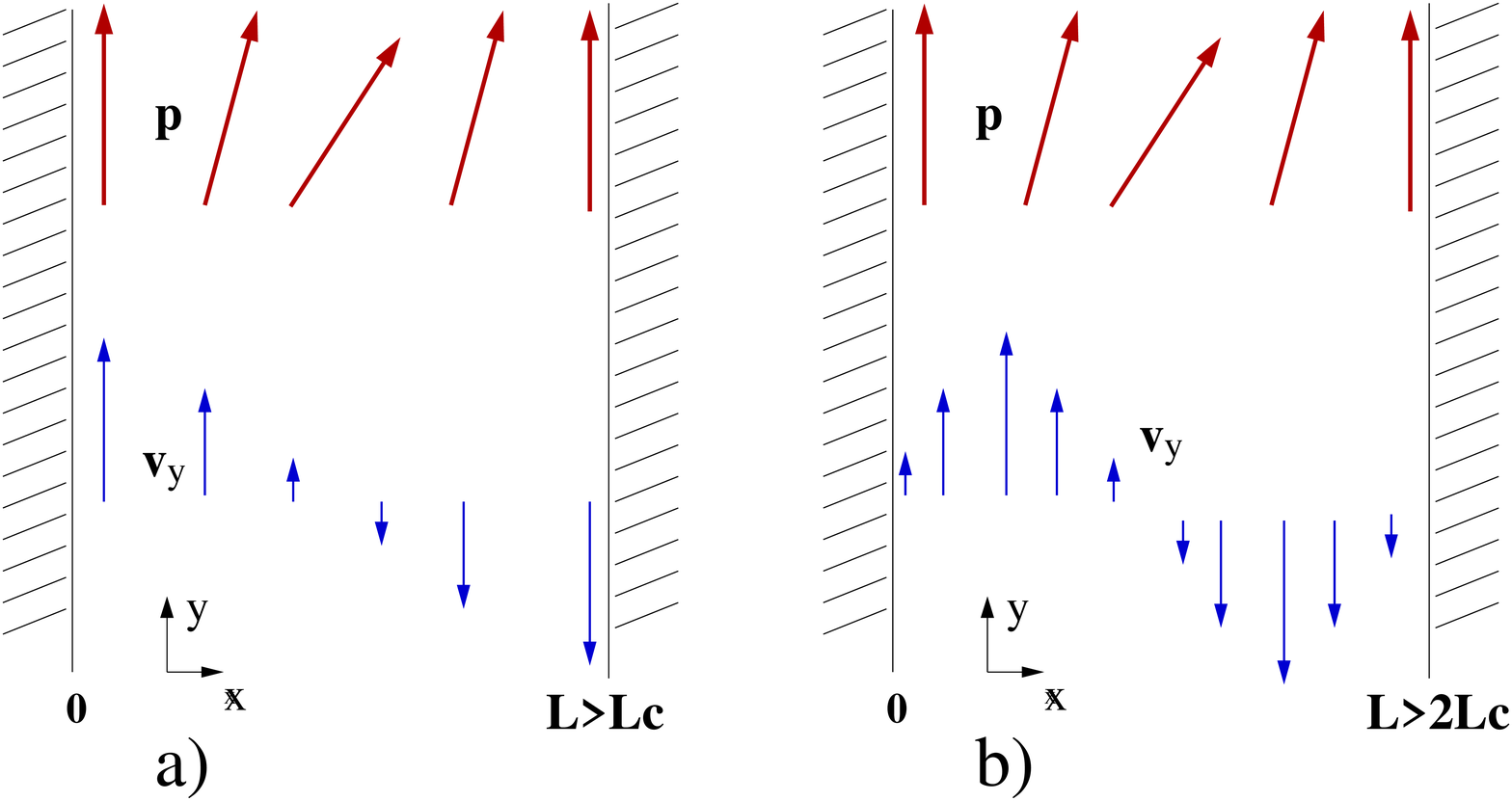}}
\caption{\label{fig3} Quasi--1d geometry: a) with free slip domain walls. The system displays a spontaneous flow transition for $L>L_c$. b)with no-slip domain walls. The system displays a spontaneous flow transition for $L>2L_c$}\end{center}
\end{figure}

In the vicinity of the spontaneous flow transition, the maximum tilt angle is small and one can again linearize the polarization equation. The polarization tilt $\epsilon$ is then given by equation \ref{tilt}. The integration of the velocity gradient with the boundary condition $v_y(x=0)=0$ leads to

\begin{equation}
v_y=\frac{4\sigma_{xy}L_c}{\pi(4\eta+\gamma(\nu+1)^2)}\left(\sin(\pi x/L_c)-\tan(\frac{\pi L}{2L_c})(\cos(\pi x/L_c)-1)\right)
\end{equation}
The boundary condition  $v_y(L)=0$ can be satisfied only if $L=2L_c$. The spontaneous flow transition occurs therefore at $L=2L_c$, or equivalently at $\Delta\mu=4\Delta\mu_c$ (see Fig\ref{fig3}b). 

Above threshold, the tilt cannot be considered as small and a full non-linear analysis is required. As in the previous case, close to the transition, the maximum tilt angle and the transverse stress $\sigma_{xy}$ scale like  $\sigma_{xy} \propto \epsilon_m\propto\sqrt{L-L_c}$.

\section{Hydrodynamics mixed boundary conditions}

\begin{figure}[h]
\begin{center}
\scalebox{0.13}{
\includegraphics{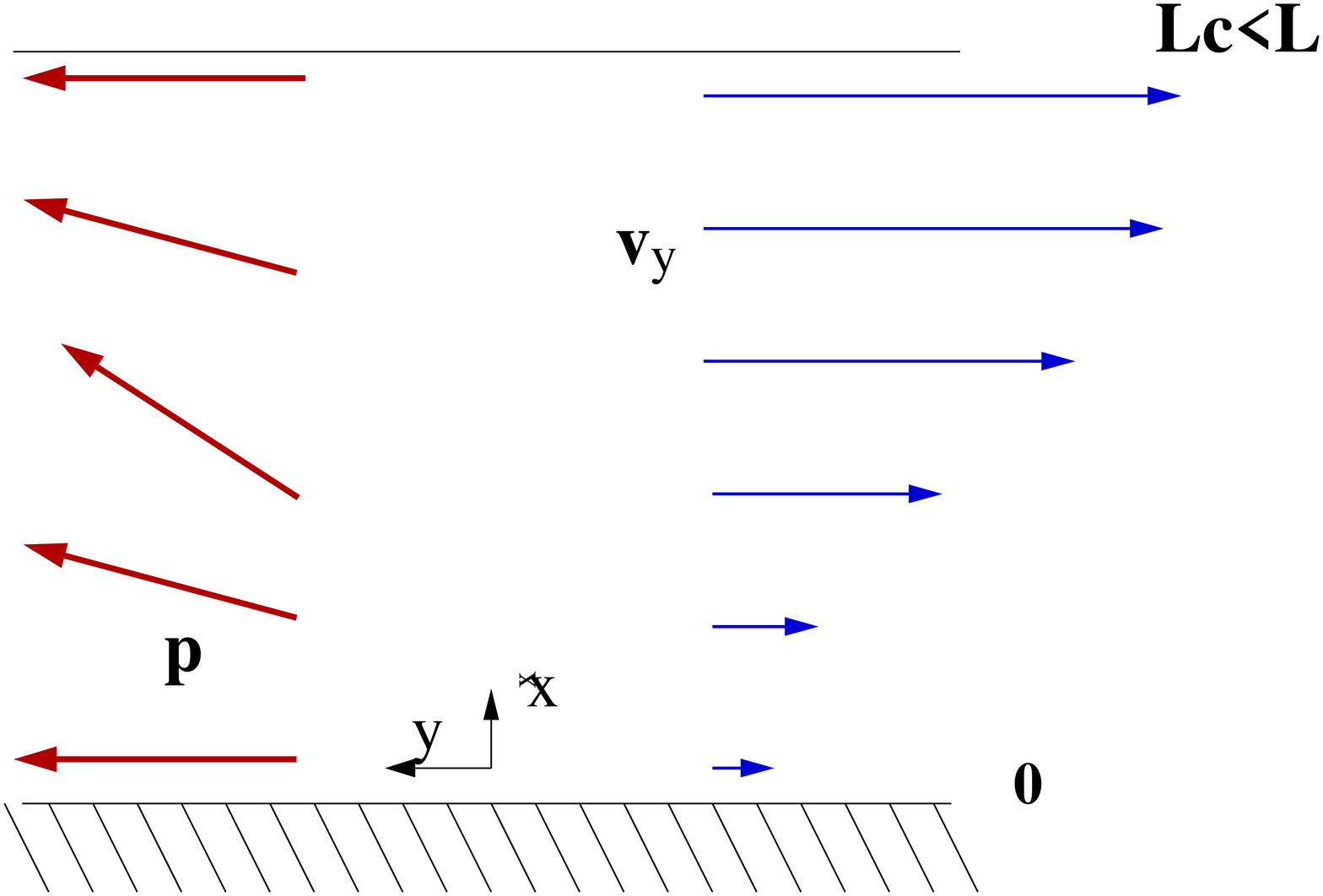}}
\end{center}
\caption{\label{fig4} A gel lying on a solid substrate and the associated flow}
\end{figure}

The easiest experimental conditions are that of a film on a solid plane with a free surface ($v_y(x=0)=0, \sigma_{xy}(x=L)=0$). As the stress vanishes everywhere, the polarization field is the same as for a free standing film with free boundary conditions. The transition threshold is given by Eq. (8). In this case a finite gel flux $Q$ is generated above threshold
\begin{equation}
Q = \int_0^L v_y dx=-\frac{4L{\tilde \zeta}\Delta\mu \quad \epsilon_m}{\pi[4\eta+\gamma(\nu+1)^2]}.
\end{equation}
with $\epsilon_{m}=-\sqrt{2\pi(L-L_c)/L_c}$. The velocity profile is sketched on figure \ref{fig4}. This provides a striking example of self-generated motion that could be observed in an annular geometry. The Fredericks transition could further play a role in lamellipodium motility \cite{albe02}. 

For completeness, we also study the case where the slip is different on the two surfaces:  $v_y(x=0)=\mu \sigma_{xy}; \quad v_y(x=L)=-\mu' \sigma_{xy}$. This could perhaps be realized experimentally by inserting 
a liquid film between the gel and a solid confining  surface and working with homeotropic boundary conditions ($\theta=0$). There is, in this case, a spontaneous flow transition at a thickness $L$ such that $L_c<L<2L_c$ given by 
\begin{equation}
\tan (\frac {\pi L}{2L_c})=-\frac{\pi(\mu+\mu')(4\eta+\gamma(\nu+1)^2)}{8L_c}
\end{equation}
If the boundary conditions are asymmetric ($\mu\neq \mu'$), the transition leads to the apparition of a finite gel flux in the direction $y$ parallel to the confining surfaces. 

\section{Active boundary conditions}

Active boundary conditions can be generated by confining the gel between two planar surfaces coated with molecular motors. The motors impose a finite gel velocity on the surfaces. If the confining surfaces are identical, the velocities on the two surfaces are equal; this is equivalent to no-slip boundary conditions with a constant drift at the motors velocity. If the two velocities imposed by the motors are different, there is always a finite shear stress $\sigma_{xy}$ and a finite tilt of the polarization: there is no continuous spontaneous flow transition in this case. 

\section{Concluding remarks}

In this letter we predict a spontaneous Frederiks--like transition in active polar materials confined in one dimension: for small thicknesses or small activity, boundary effects are prevailing and the gel remains in an unperturbed, static,  homogeneous state. Above a critical thickness or a critical activity, a polarization tilt appears and the system flows.  The transition depends strongly on the very nature of the boundary conditions imposed by the confinement: homeotropic or planar anchoring are necessary, and the threshold value depends on the nature of the hydrodynamic boundary conditions. Interestingly, for asymmetric active boundary conditions, there is no continuous transition.

A detailed experimental study of the transition, using for instance actin-myosin gels in micro-channels is very promising. A measurement of the transition threshold 
will give direct access to the effective active stress $\tilde \zeta \Delta\mu$. 
The two-fold aspect of the transition, involving both dynamical and polarization 
properties should allow for various methods of visualization coming either from liquid crystal physics or from micro-fluidics. We are not aware at the moment of any 
quantitative measurement of these active stresses that would be essential 
to give a proper mesoscopic description of active actin gels. More generally, 
any geometry where anchoring imposes a non uniform polarization field in 
an active gel would lead to a hydrodynamic flow driven by the activity because 
of the couplings between polarization, activity and flow. A careful choice of 
the geometry and of the anchoring conditions could then allow for a 
determination of the various active coefficients of the gel ($\zeta$, $\lambda$) 
as well as that of the more standard liquid crystalline parameters ($\nu$,$\gamma$). 

Our results are remarkable in the sense that they predict a 
spontaneous flow transition in the absence of any external field. This opens the way to a quantitative characterization of this new class of materials. We think 
that the understanding of this physics 
will prove to be useful in the study of the cell cytoskeleton. In particular,  lamellipodia share several features with a polar active gel slab.  

\acknowledgments
We are grateful to S.Fraden (Boston) for important discussions 
on possible experimental realizations of the spontaneous flow transition in active polar gels.


\begin{thebibliography}{0}
\bibitem{albe02}
B. Alberts {\it et al.},
{\it Molecular Biology of the Cell}  4th ed. (Garland, New York, 2002).

\bibitem{howa01}
J. Howard,
{\it Mechanics of Motor Proteins and the Cytoskeleton} (Sinauer Associates, Inc., Sunderland, 2001).

\bibitem{nede97}
F.J. N\'ed\'elec {\it et al.},
{\it Nature}  {\bf 389}, 305 (1997),
F.J. N\'ed\'elec, T. Surrey, and A.C. Maggs,
{\it Phys. Rev. Lett.} {\bf 86}, 3192 (2001), T. Surrey {\it et al.},
{\it Science} {\bf 292}, 1167 (2001).


\bibitem{krus00}
K. Kruse and F. J\"ulicher,
{\it Phys. Rev. Lett.} {\bf 85}, 1778 (2000).

\bibitem{krus01}
K. Kruse, S. Camalet, and F. J\"ulicher,
{\it Phys. Rev. Lett.} {\bf 87}, 138101 (2001).
\bibitem{lee01}
H.Y. Lee and M. Kardar,
{\it Phys. Rev. E} {\bf 64}, 056113 (2001).

\bibitem{kim03}
J. Kim {\it et al.},
{\it J. Korean Phys. Soc.} {\bf 42}, 162 (2003).

\bibitem{menon}
S. Sankararaman, G.I. Menon and P.B. Sunil Kumar,
{\it Phys. Rev. E} {\bf 70}, 031905 (2004),
M.C. Aronson {\it Unpublished}

\bibitem{live03}
T.B. Liverpool and M.C. Marchetti,
{\it Phys. Rev. Lett.} {\bf 90}, 138102 (2003).

\bibitem{simh02}
A. Simha and S. Ramaswamy, {\it Phys. Rev. Lett.} {\bf 89}, 058101 (2002).

\bibitem{warner}E.M. Terentjev, M. Warner, R.B. Meyer and J. Yamamoto, Phys. Rev.E {\bf 60} (1999)

\bibitem{seki91}K.Sekimoto J:Phys.II (France) {\bf 1}, 19 (1991).

\bibitem{thou} O. Thoumine, A. Ott, {\it J. Cell. Sc.} {\bf 110} (1997)
\bibitem{uhde04} J. Uhde, M. Keller and E. Sackmann, {\it Phys. Rev. Lett.} {\bf 93} (2004)
\bibitem{krus04} K. Kruse, J.F. Joanny, F. J\"ulicher, J. Prost and K. Sekimoto,
Phys. Rev. Lett {\bf 92}, 078101 (2004). 

\bibitem{krus05} K. Kruse, J.F. Joanny, F. J\"ulicher, J. Prost and K. Sekimoto,
Eur. Phys. J. E {\bf 16}, 5 (2005). 

\bibitem{head03}
D.A. Head, A.J. Levine, F.C. MacKintosh, {\it Phys. Rev. Lett.} {\bf 91}, 108102 (2003).

\bibitem{wilh03}
J. Wilhelm and E, Frey, {\it Phys. Rev. Lett.} {\bf 91}, 108103 (2003).

\bibitem{taki91} K.Takiguchi J.Biochem. {\bf 109}, 520 (1991).



\bibitem{dege93}
P.G. De Gennes and J. Prost,
{\it The Physics of Liquid Crystals} (Clarendon Press, Oxford 1993).



\end{thebibliography}
\end{document}